\documentclass[conference, compsoc]{IEEEtran}
\usepackage[nocompress]{cite}
\usepackage[pdftex]{graphicx}
\usepackage{amsmath}
\usepackage{array}
\usepackage{url}

\usepackage{amsthm}

\usepackage{algorithm, algorithmicx}
\usepackage[noend]{algpseudocode}

\usepackage{enumitem}
\usepackage{booktabs}

\usepackage{etoolbox, siunitx}
\robustify\bfseries
\usepackage{xspace}
\usepackage[usenames,dvipsnames]{xcolor}

\newcommand{\WCTruss}{$WC$\xspace}
\newcommand{\rkt}{$Ros$\xspace}
\newcommand{\pkt}{\textsc{PKT}\xspace}

\newcommand{\ktruss}{$k$-truss\xspace}
\newcommand{\ktrusses}{$k$-trusses\xspace}
\newcommand{\kclass}{$k$-class\xspace}

\newcommand{\TMax}{$t_\text{max}$\xspace}

\newcommand{\eidA}{$\mathit{Eid}$\xspace}
\newcommand{\supA}{$\mathit{S}$}
\newcommand{\adj}{$\mathit{N}$\xspace}
\newcommand{\numEdges}{$\mathit{Es}$\xspace}
\newcommand{\edgeStart}{$\mathit{Eo}$\xspace}
\newcommand{\edgeList}{$\mathit{El}$\xspace}

\newcommand{\X}{$\mathit{X}$}

\newcommand{\parTri}{\textsc{ParTriangle-Ros}\xspace}
\newcommand{\parTriAM}{\textsc{ParTriangle-AM4}\xspace}
\newcommand{\inCurr}{$\mathit{inCurr}$\xspace}
\newcommand{\inNextA}{$\mathit{inNext}$\xspace}
\newcommand{\processedA}{$\mathit{processed}$\xspace}

\newcommand{\kcore}{$k$-core\xspace}
\newcommand{\buff}{$\mathit{buff}$\xspace}
\newcommand{\curr}{$\mathit{curr}$\xspace}
\newcommand{\nextA}{$\mathit{next}$\xspace}

\newcommand{\ourpark}{\textsc{PKC}\xspace}

\newcommand{\park}{ParK\xspace}

\newcommand{\mpm}{MPM\xspace}

\newcommand{\scan}{\textsc{Scan}\xspace}
\newcommand{\procsublvl}{\textsc{ProcessSubLevel}\xspace}

\begin{document}

\title{Shared-memory Graph Truss Decomposition}
\author{\IEEEauthorblockN{Humayun Kabir~~~~Kamesh Madduri}
\IEEEauthorblockA{Computer Science and Engineering\\
The Pennsylvania State University\\
University Park, PA, USA\\
Email: \{hzk134, madduri\}@cse.psu.edu}
}
\maketitle

\begin{abstract} 

We present PKT, a new shared-memory parallel algorithm and OpenMP implementation for the truss decomposition of large sparse graphs. A \ktruss is a dense subgraph definition that can be considered a relaxation of a clique. Truss decomposition refers to a partitioning of all the edges in the graph based on their \ktruss membership. The truss decomposition of a graph has many applications. We show that our new approach PKT consistently outperforms other truss decomposition approaches for a collection of large sparse graphs and on a 24-core shared-memory server. PKT is based on a recently proposed algorithm for \kcore decomposition.
\end{abstract}



\section{Introduction}
\label{s:intro}
Graphs are ubiquitous. Any set of interacting entities can be represented as a graph, with the entities as vertices and interactions as edges. To understand the structure of a graph, it is often useful to find densely connected sets of vertices and edges, or cohesive subgraphs, in the graph. There are many well-known notions of cohesive subgraphs. A maximal clique is likely the oldest and most used definition for a cohesive subgraph. Several relaxations of a clique have also been proposed. For instance, an \mbox{$n$-clique}~\cite{Luce1950} relaxes the distance between any two vertices in a clique to be $n$ instead of 1. An $n$-clan~\cite{Mokken:1979} is defined as an $n$-clique such that the diameter is bounded by $n$, and an $n$-club~\cite{Mokken:1979} is a maximal subgraph of diameter $n$. A 
$k$-plex~\cite{doi:10.1080/0022250X.1978.9989883} relaxes the internal connectivity of vertices in a clique, with a 1-plex being a clique.  Quasi-clique formulations (e.g., \cite{Abello2002, Pei:2005:MCQ:1081870.1081898}) relax other constraints, such as the subgraph edge density or vertex degrees. However, since most clique-based problem formulations are NP-complete, exact computation is computationally intensive for large graphs.

\begin{figure}[!t]
\centering  
\includegraphics[width=0.48\textwidth]{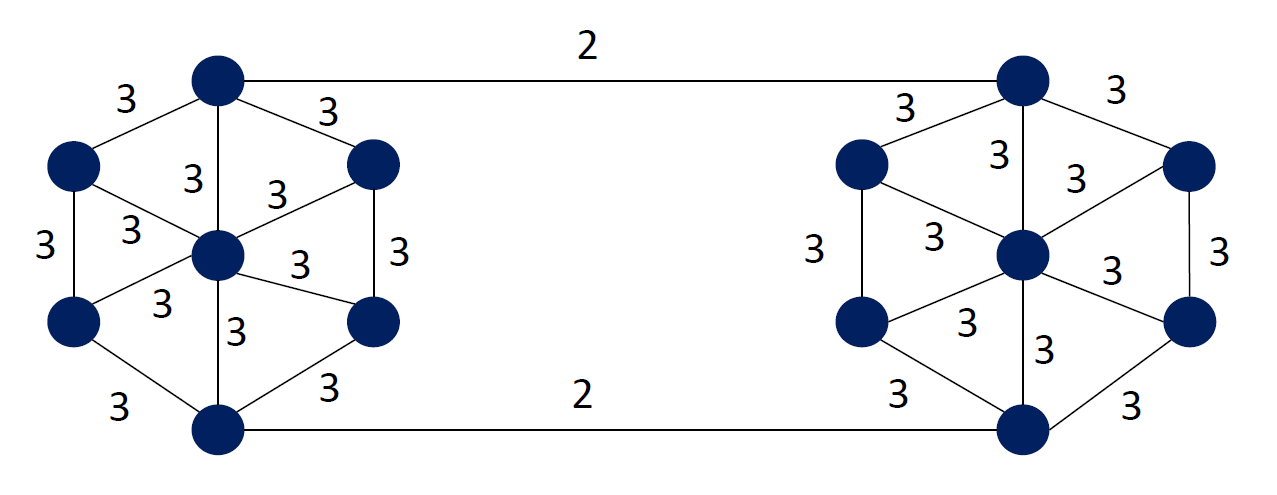}
\caption{An example graph showing \kcore and \ktruss decomposition. All vertices have a coreness value of 3. Two edges have trussness of 2 and the rest of the edges have trussness value of 3. There are two 3-trusses in this graph.}   
\label{fig:ktruss}
\end{figure}

The \kcore~\cite{Sei83, MB83} and \ktruss~\cite{CohenTruss} cohesive subgraph formulations are very useful in practice because they can be computed exactly using simple polynomial-time algorithms. Both these formulations can also be used for a hierarchical decomposition of the graph. A \kcore is a maximal subgraph such that each vertex has degree at least $k$. A \ktruss~\cite{CohenTruss} is defined as a maximal non-trivial single-component subgraph such that every edge is contained in at least $k-2$ triangles. A vertex is said to have a coreness value or coreness of $k$ if it belongs to a $k$-core of the graph, but not the $(k+1)$-core. Similarly, an edge is defined to have a trussness of $l$ if it belongs to an $l$-truss of the graph, but not the $(l+1)$-truss. As Cohen writes in the paper introducing \ktrusses~\cite{CohenTruss}, ``a \ktruss provides a nice compromise between the too-promiscuous ($k-$1)-core and the too-strict clique of order $k$.'' The problem of \ktruss decomposition refers to computing the trussness value of every edge in the graph. \kcore decomposition is similarly defined. Figure~\ref{fig:ktruss} shows trussness values of all the edges in a simple undirected graph. Given edge trussness values, the maximal \ktruss subgraphs (for a specific $k$) can be determined by executing connected components on the graph after deleting edges with trussness less than $k$. A \ktruss is (almost) identical to the k-dense~\cite{SYK06}, triangle $k$-core~\cite{6228155}, and the k-community~\cite{VB13} cohesive subgraph formulations independently presented by other authors. Sar{\i}y{\"u}ce et~al.~\cite{SSPC15} recently presented a new formulation for cohesive subgraphs called nucleus decompositions, which subsumes both \kcore and \ktruss definitions. 

\ktruss decomposition has numerous uses in large-scale graph analysis, including visualization~\cite{DBLP:conf/nips/Alvarez-HamelinDBV05}, preprocessing for community detection~\cite{SYK06, VB13, Huang:2015:ACC:2856318.2856323} and maximal clique finding~\cite{DBLP:journals/corr/abs-1302-6256}. There are also several sequential and parallel algorithms for \ktruss decomposition~\cite{CohenTruss, Wang:2012:TDM:2311906.2311909,6228155,5076317, 7004264,SSP17,Graphulo}, which we will introduce in the next section. Truss decomposition is also part of a recent graph benchmarking effort~\cite{GraphChallenge}. 

In this paper, we present \pkt, a new shared-memory parallel algorithm for truss decomposition. The following are the key features of \pkt and the new contributions of this work:
\begin{itemize}[leftmargin=*]
\item We perform a level-synchronous parallelization of the best sequential algorithm for \ktruss decomposition~\cite{Wang:2012:TDM:2311906.2311909}. The parallelization is similar to \park~\cite{7004366}, an algorithm for \kcore decomposition. We avoid the inherently-sequential edge processing approach in the sequential algorithm.  
\item Our approach is memory-efficient in that the memory requirements are proportional to the number of edges and not the number of triangles in the graph.
\item Unlike other \kcore and \ktruss algorithms, we do not use a hash table to maintain edges in the graph. Instead, we use data structures that are amenable to safe and easy concurrent updates.
\item For support computation, a key subroutine of many truss decomposition approaches, we use a highly tuned and graph ordering-aware method that performs significantly lower work for graphs with skewed degree distributions. 
\item Our algorithm is work-efficient for most real-world graphs.
\end{itemize}


\section{Background}
\label{s:background}



Let $G = (V, E)$ be an undirected and simple graph with one connected component, $n = |V|$ vertices, and $m = |E|$ edges. We use $N(u)$ or $Adj(u)$ to denote the set of neighbors of a vertex $u$, i.e., $N(u) = \{v : \langle u,v\rangle \in E\}$.  The degree of a vertex $u$ is denoted by $d(u) = |N(u)|$. A triangle in $G$ is a cycle of length 3. We denote a triangle by the edge triple that forms it: $\langle u, v\rangle$, $\langle v, w\rangle$, and $\langle u, w\rangle$. The order of $u$, $v$, and $w$ does not matter when denoting the triangle. The set of all triangles in $G$ are denoted by $\triangle_G$. Similarly, we define a wedge to be a pair of edges with a common endpoint, e.g., $\langle u, v\rangle$ and $\langle v, w\rangle$. Triangles can be viewed as closed wedges, i.e., the edge $\langle u, w\rangle$ is present. Let $\bigwedge$ denote the set of wedges in the graph. We define support of an edge $e = \langle u,v\rangle \in G$, $S(e, G)$, as the number of triangles that it is contained in. A \ktruss~\cite{CohenTruss, 5076317} is then defined as follows: A \ktruss $T_k$ ($k \geq 2$) is a maximal connected subgraph of $G$ such that for each edge $e \in T_k$, $S(e, T_k) \geq k-2$. We use $t_\text{max}$ to denote the maximum trussness of any edge in $G$. The \kclass of $G$ is the set of all edges with trussness $k$. The truss decomposition problem refers to finding the trussness of every edge.

In the paper introducing the \ktruss subgraph, Cohen also gives an algorithm for enumerating maximal trusses. To list \ktrusses for a specific $k$, this algorithm first computes the support of each edge in $G$. Next, it finds the edges with support less than $k-2$ and removes them. When removing an edge, the support of the edges that form a triangle with the removed edge are reduced. In Cohen's algorithm, for each edge $e$ = $\langle u,v\rangle$, computing its support and processing it takes time proportional to $d(u) + d(v)$. The total time for this algorithm is thus $\Theta(n + m + \sum_{\langle u,v\rangle \in E} (d(u) + d(v))$, which simplifies to $\Theta(n + m + 2 \sum_{v \in V} d(v) ^ 2)) = \text{O}(m^{1.5})$, since $m = \text{O}(n^2)$ for a simple graph. Cohen also proposed a MapReduce algorithm for computing \ktrusses~\cite{5076317}.

Wang et al.~\cite{Wang:2012:TDM:2311906.2311909} define the problem of \ktruss decomposition and present an algorithm for computing the trussness of every edge. We call this algorithm \WCTruss and its steps are given in Algorithm~\ref{alg:serialtruss}. Like Cohen's algorithm, this approach starts by computing the support of each edge. Next, the edges are sorted in ascending order of their support using a linear-time sort such as counting sort. Edges are then processed in increasing order of support.  Each edge is processed exactly once, and for an edge $e = \langle u,v\rangle$, a \emph{canonical} edge representation assuming $d(u) \leq d(v)$ is used. For each neighbor $w$ of $u$, the algorithm checks if $u$, $v$, and $w$ form a triangle or not.  This is done by using a hash table, where the keys are a pair of vertices. Given a pair of vertices, the hash table checks if the pair is a graph edge or not. If  $u, v, w$ form a triangle $\triangle_{uvw}$, the support of edges $\langle u,w\rangle$ and $\langle v,w\rangle$ are decreased, if their support is greater than the support of $e$. The edges are then reordered according to their new support. The edge $e$ is removed from the hash table when all the triangles containing $e$ have been processed. The operation counts for steps 5 to 16 is again bounded by  $\sum_{v \in V} d(v) ^ 2$, assuming hash table lookup and delete operations are constant time and assuming that an edge reordering can also be performed in constant time. The technique to accomplish constant time reordering is very similar to the one used in the \kcore algorithm by Batagelj and Zaversnik~\cite{DBLP:journals/corr/cs-DS-0310049}. The support of all the edges can be computed in $\text{O}(m^{1.5})$ time, and so the overall time complexity of this decomposition algorithm is $\text{O}(m^{1.5})$. This algorithm has two disadvantages: step 6 makes it inherently sequential, and hash table operations can be expensive in practice. \WCTruss processes all the edges belonging to a \kclass before processing edges belonging to $(k+1)$-class, and so this approach can be considered a bottom-up strategy.  Also, note that when an edge is processed, the edges forming a triangle with that edge may also become part of the current \kclass, as their support is decreased. The initially computed support for an edge $e$ (or more precisely, $S[e]+2$) is an upper bound for the eventual trussness of the edge. The Graphulo algorithm for \ktruss decomposition~\cite{Graphulo} is based on Wang~et~al.'s approach, but uses linear algebra primitives and array-based building blocks. 


\begin{algorithm}[!t]
  \caption{\WCTruss: Serial \ktruss decomposition algorithm.}
  \label{alg:serialtruss}
  \begin{algorithmic}[1]
   
   \Statex \Procedure{\ktruss-\WCTruss}{$G$, \supA}
  
  \State Compute support $S[e]$ for all $e \in E$.  
  \State {Using a $\Theta(m)$-time sort, order edges by support 
  \Statex ~~~~~and store them in $El$}  
  \State {Add all $e \in E$ to a hash table $Eh$}
  \While {$El \neq \phi$}
  	\State Extract $e=\langle u,v\rangle$, the edge with the lowest 
    \Statex ~~~~~~~~~support from $El$
  	\State $k = S[e]$  	
  	\For{($w \in N(u)$)}
  	\If{$\langle v,w\rangle$ $\in Eh$} 
  	\If{$S[\langle u,w\rangle] > k$} 
  		  \State {$S[ \langle u,w\rangle] \gets S[\langle u,w\rangle] - 1$}  
   		  \State {Reorder $El$}  
  	\EndIf  	

  	\If{$S[ \langle v,w\rangle] > k$} 
  	  		  \State {$S[\langle v,w\rangle] \gets S[\langle v,w\rangle] - 1$} 
   		  \State {Reorder $El$}  
  	\EndIf    	
  	
  	\EndIf
  	
  	\EndFor
  	
  	\State Remove $e$ from $Eh$
  	
    \EndWhile
    
    \State Increment all entries of $S$ by 2 to get trussness.
      
    \EndProcedure  
  
	\end{algorithmic}
\end{algorithm}


Wang et al.~\cite{Wang:2012:TDM:2311906.2311909} also propose two external-memory algorithms, a  bottom-up algorithm and a top-down algorithm.  In the bottom-up algorithm, the graph is divided into $p$ parts such that each part can fit in main memory. The algorithm starts by computing a lower bound of \ktruss values for each edge.  It then lists all $T_k$, $2 \leq k \leq t_\text{max}$.  To find a \ktruss, the algorithm forms a graph with vertices that are end points of edges $e$, such that the trussness of these edges is at least $k$.  The \ktruss subgraph is found using this constructed graph and the algorithm continues until all the $k$-trusses are listed. The top-down approach computes a trussness upper bound for each edge.  Given these bounds, it finds the maximum of all the upper bounds of truss numbers and constructs a graph using the edges corresponding to this maximum upper bound of all edges. It uses this constructed graph to find the \ktrusses corresponding to the $t_\text{max}$-class. The authors observe that the top-down approach is preferable if we only want to list trusses for large $k$. 

Zhang and Parthasarathy~\cite{6228155} introduce the triangle \kcore formulation, which is almost identical to a \ktruss. Their algorithm is different in that all triangles are listed in the support computation step, and the data structure storing them is updated in the subsequent steps. The space requirements of this algorithm scale as $\text{O}(|\triangle|)$, but superfluous triangle lookups are avoided and a hash table is not required. We do not consider parallelization of this algorithm because of the considerably higher space requirements.

\begin{algorithm}[!t]
  \caption{Parallel triangle counting.}
  \label{alg:parsupport}
  \begin{algorithmic}[1]
  
  	\Procedure{\parTri}{$G$, \supA, \X}
 
  \For{$\langle u,v\rangle$ $\in E$ \textbf{in parallel} }

 	 \For{($w \in N(u)$)}
  		\X[w] = $\langle u,v\rangle$
	\EndFor

	\For{($w \in N(v)$)}
  		\If{$w \neq u$} 
			\If{\X$[w]$ = $\langle u,v\rangle$} 
				\State {\supA($\langle u,v\rangle$) $\gets$ \supA($\langle u,v\rangle$) $+ 1$} 
			\EndIf
  		\EndIf  	

	\EndFor
\EndFor 

	\EndProcedure
  
	\end{algorithmic}
\end{algorithm}

\begin{figure}[!t]
\centering  
\includegraphics[width=0.48\textwidth]{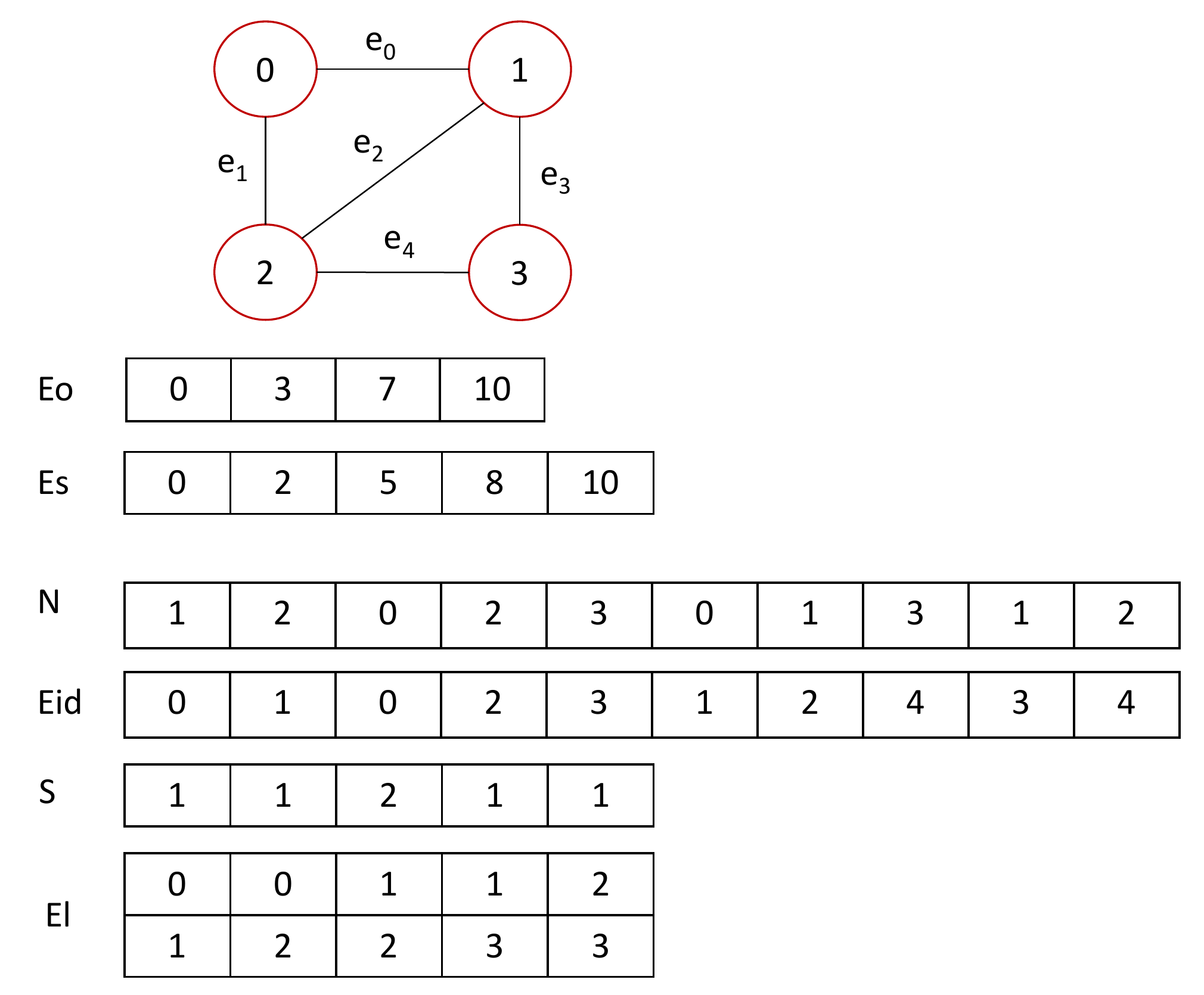}
\caption{Graph representation to compute \ktruss. An array eid is used to store the edge id corresponding to an adjacency of a vertex. Here, $n = 4$, and $m = 5$.}   
\label{fig:CSRAugment}
\end{figure}

Rossi~\cite{Rossi2014} presents an algorithm for truss decomposition that parallelizes just the support computation phase. He augments the compressed sparse row representation of a graph with an extra array \eidA to store edge identifiers corresponding to each neighbor of a vertex.  An example of this representation for a small graph is shown in Figure~\ref{fig:CSRAugment}. The data structure uses an edge list \edgeList to store the vertex tuples. This representation helps compute trusses without using a hash table. Rossi's parallel support computation approach is outlined in Algorithm~\ref{alg:parsupport}.  The algorithm uses an edge-based approach to count triangles in parallel.  Each thread initializes a temporary array \X of size $n$.  To compute support of an edge $\langle u,v\rangle$, the neighbors of $u$ are first marked in the array \X.  The thread then visits the neighbors $w$ of $v$ and checks if \X$[w]$ is marked or not. If the entry is marked, then  $u,v,w$ forms a triangle $\triangle_{uvw}$, and so the support of  $\langle u,v\rangle$ is incremented by 1.  
For each edge $e$ = $\langle u,v\rangle$, the support computation takes time proportional to $d(u) + d(v)$.  Thus, the time for this algorithm is proportional to $\sum_{e=\langle u,v\rangle} (d(u) + d(v)) = 2 \sum_{v} d(v) ^ 2 $. 
Subsequent steps in Rossi's algorithm are similar to Algorithm~\ref{alg:serialtruss}.  We refer to this algorithm as \rkt in the rest of the paper. 

Chen et al.~\cite{7004264} propose a distributed algorithm for \ktruss decomposition. They provide a better MapReduce algorithm for truss decomposition compared to~\cite{5076317}. They also develop a distributed algorithm for \ktruss decomposition using Pregel~\cite{Malewicz:2010:PSL:1807167.1807184}.

Edge support computation is closely related to the problems of exact triangle counting and listing. Compared to truss decomposition, triangle counting is a very well-studied problem. We refer readers to~\cite{CN85,Lat08, OB14,shun7113280, CXL16,Xiao:2017:ACT:3034786.3034790,parimalarangan_parsocial2017} for a sampling of efficient algorithms and practical high-performance implementations. Xiao~et~al.~\cite{Xiao:2017:ACT:3034786.3034790} unify a large body of previously-developed triangle counting algorithms and observe that the ordering of vertices and orientation of edges has a significant impact on performance. We use a parallel triangle counting implementation proposed for the problem of triad census in directed graphs ~\cite{parimalarangan_parsocial2017}. 

\kcore decomposition is closely related to truss decomposition. As mentioned previously, the \WCTruss algorithm is very similar in design to Batagelj's and Zaversnik's linear-time \kcore decomposition algorithm~\cite{DBLP:journals/corr/cs-DS-0310049} (referred to as the BZ algorithm). Cohen also proves a number of properties relating \kcore and \ktruss subgraphs. A key \ktruss motivation is that it helps find vertex clusters that are more cohesive than a typical \kcore. 

The parallel \kcore decomposition algorithm \park~\cite{7004366} can be considered a  level-synchronous parallelization of the BZ algorithm.  In \park, the coreness of all vertices are computed in a bottom-up manner.  The algorithm uses two arrays \curr and \nextA to store vertices in the current and next level.  The algorithm scans an array that contains the current estimate of the coreness values and puts the vertices in \curr array.  It then processes the vertices in the \curr array.  This may decrease the coreness values of neighbor vertices and may put them in current level, and such vertices are added to the \nextA array.  The arrays are swapped at the end of each phase, and this process goes on until all the vertices are processed.  If $c_\text{max}$ is the maximum coreness value in the graph, the work performed by this  algorithm is $\text{O}(nc_\text{max}+m)$. We further improved this algorithm~\cite{kabir_parsocial2017}.

An alternate algorithm for \kcore decomposition was proposed by Montresor, De Pellegrini, and Miorandi~\cite{DBLP:journals/corr/abs-1103-5320}. We refer to this approach as the \mpm algorithm. This algorithm uses a simple local update rule that is repeatedly applied at every vertex. The authors show that when starting with the degrees and applying this update rule for several iterations, the degrees converge to coreness values. While this approach is not work-efficient (since each edge is processed multiple times), an advantage is that it does not require fine-grained synchronization. For this reason, MPM can be easily adapted to distributed settings.  Sar{\i}y{\"u}ce~et~al.~\cite{SSP17} recently proposed shared-memory parallelization of their nucleus decomposition formulation based using the update rule of the \mpm algorithm. Since a \ktruss is a special case of their nucleus decomposition, the  Sar{\i}y{\"u}ce~et~al. local algorithm can be considered an alternative to the level-synchronous parallelization strategy.



\section{Parallel \ktruss Decomposition}
\label{s:algorithms}

We present our algorithm \pkt in this section.  We begin by describing the support computation approach, which is in turn based on a recent parallel triangle counting algorithm~\cite{parimalarangan_parsocial2017}.    

The data structures used to store the graph are illustrated in Figure~\ref{fig:CSRAugment}.  In addition to the compressed sparse row representation (\adj, \numEdges), four arrays are used.  An array \eidA of size $2m$ is used to store the edge id corresponding to each neighbor of a vertex. An array $S$ of size $m$ is used to store the support of each edge. An array \edgeStart of size $n$ is used to store the index of first neighbor greater than a vertex. Finally, an array \edgeList of size $m$ is used to store the edge list, i.e., the vertices corresponding to each edge. Thus, assuming 4-byte integers, the space requirement is $(n + 2m + 2m + m + n + 2m)\times 4$ bytes $= 28m+8n$~bytes.


\noindent \textbf{Parallel support computation.} Fast algorithms for triangle counting use degree- or \kcore-based vertex ordering and combine this with edge orientation. With increasing \kcore-ordering of vertices, a canonical triangle representation of $u < v < w$ gives a low operation count~\cite{OB14,Xiao:2017:ACT:3034786.3034790}. We define $N^+(u) = \{v : \langle v, u \rangle \in G , v > u\}$, $d^+(u) = | N^+(u) |$, $N^-(u) = \{v : \langle v, u \rangle \in G , v < u\}$, and $d^-(u) = | N^-(u) |$. 

Our parallel support computation is given in Algorithm~\ref{alg:parsupportAM4}. For every vertex $u$, we mark the neighbors $N^+(u)$  using a thread-local array \X.  We then visit each $v \in N^-(u)$ and consider adjacencies $w \in N^+(v)$.  If $w$ is marked, $vuw$ forms a triangle $\triangle_{wuv}$ and $v < u < w$.   The time complexity of this algorithm is  $\Theta(\sum_v ( d^+(v) + d^+(v) ^ 2 ) ) = \Theta(m + \sum_v (d^+(v) ^ 2))$.

In triangle counting, the array $X$ could be a bit vector. However, for support computation, we use $X$ to store the edge id of the edge $\langle u, w \rangle$.  Also, no atomic operations are necessary in triangle counting. However, three atomic operations are used for each triangle discovered, to count the support of each edge in the triangle.  This adds overhead to  support computation that is not present in triangle counting. 

\begin{algorithm}[!t]
  \caption{AM4: Parallel triangle counting.}
  \label{alg:parsupportAM4}
  \begin{algorithmic}[1]
  
  	\Procedure{\parTriAM}{$G$, \supA, \X}
	\State {$X \gets \phi$} \Comment{thread-local temp array} 

 \For{($u = 0$ to $n -1$) \textbf{in parallel}}
 	
 	\For{($j = Eo[u]$ to $Es[u+1] -1$)  }
    	\State $w \gets N[j]$
    	\State{$X[w] \gets j+1$}
    \EndFor

 	\For{ ($j = Es[u]$ to $Eo[u]-1$) }
    	\State $v \gets N[j]$
        \For{ ($k = Es[v+1] -1$ down to $Eo[v]$) }
        	\State $w \gets N[k]$
        	\If {($w < u$)}
            	\State break
            \EndIf
            
        	\State $e_{vw} \gets Eid[k]$
            \State $e_{vu} \gets Eid[j]$
            \State $e_{uw} \gets Eid[ X[w] - 1]$
            
            \State AtomicAdd(\supA $[ e_{vw} ]$, 1)
            \State AtomicAdd(\supA $[ e_{vu} ]$, 1)
            \State AtomicAdd(\supA $[ e_{uw} ]$, 1)

        \EndFor

    \EndFor
    
    \For{($j = Eo[u]$ to $Es[u+1] -1$)  }
    	\State $w \gets N[j]$
    	\State{$X[w] \gets 0$}
    \EndFor

 \EndFor

	\EndProcedure
  
	\end{algorithmic}
\end{algorithm}



\begin{algorithm}[!t]
  \caption{Parallel \ktruss algorithm.}
  \label{alg:pktdescription}
  \begin{algorithmic}[1]
	\Procedure{\pkt}{$G$, \supA}
		\State {Initialize thread-local array \X}  
		\State {\parTriAM{(}$G$, \supA, \X{)} }
        \State {\curr $\gets \phi$; \nextA $\gets \phi$} 
		\State {\inCurr $\gets \phi$; \inNextA $\gets \phi$}
		\State {\processedA $\gets \phi$}
			
		\State {$\mathit{todo} \gets n$; $l \gets 0$} 
		\While {$\mathit{todo} > 0$} \label{pkt:start}
			\State {\scan(\supA, $l$, \curr, \inCurr)} 
			\While{$|\mathit{curr}| > 0$}
				\State {$\mathit{todo} \gets \mathit{todo} - |\mathit{curr}|$}
				\State {{\procsublvl}(\curr, \supA, $l$, 
                \Statex ~~~~~~~~~~~~~\nextA, \inCurr,\inNextA, \processedA, \X)} 
				
				\State {\curr $\gets$ \nextA} \label{serial:start}
				\State {\nextA $\gets \phi$} 
				\State {\inCurr $\gets$ \inNextA} 
				\State {\inNextA $\gets \phi$}  \label{serial:end}

			\EndWhile

			\State {$l \gets l + 1$} \label{pkt:end}
		\EndWhile	
	\EndProcedure
~~\\
  \Procedure{\scan}{\supA, $l$, \curr, \inCurr} 
    \State {Initialize a thread-local array \buff of size $s$} 
    \State {$i \gets 0$} \Comment{thread-local variable}
    \For{($e = 0$ to $m-1$) \textbf{in parallel}}
	   \If{(\supA$[e] = l)$} 
		  \State {$\mathit{buff}[i] \gets e$}; $i \gets i + 1$
		  \State {\inCurr$[e] \gets true$}
		  \If {($i = s$)}
			 \State Atomically update end of \curr 	
			 \State {Copy \buff to \curr}
			 \State \buff $\gets \phi$; {$i \gets 0$} 			
		  \EndIf
	   \EndIf
    \EndFor
    \If {($i > 0$)}
	  \State Atomically update end of \curr 	
      \State {Copy \buff to \curr}
	  \State \buff $\gets \phi$; $i \gets 0$
    \EndIf
  \EndProcedure
	\end{algorithmic}
\end{algorithm}


\textbf{Level-synchronous parallelization.} In \pkt, we compute trussness values in a bottom-up manner. We process edges belonging to the $l$-class before processing edges belonging to $(l+1)$-class. The steps are given in Algorithm~\ref{alg:pktdescription}.  The overall strategy is similar to the \park~\cite{7004366} algorithm for parallel \kcore decomposition. The output is the array $S$ indexed by the edge id containing the updated support of all the edges. Thus, the trussness of an edge $e$ is \supA$[e]+2$.   

The algorithm starts by computing the support of the edges in parallel using the previously discussed approach, and stores the support in the array \supA.  Next, it uses the procedures \scan and \procsublvl to find edges in a \kclass.  To find edges in a \kclass, the algorithm uses a \scan phase to scan the $S$ array and find edges with support $k-2$.  These edges are processed in the procedure \procsublvl. The processing of these edges in \procsublvl may add new edges to the \kclass. This continues until no more edges can be added to \kclass.

We use an array \processedA to mark if an edge has been processed or not (i.e., we mark to represent deleted edges). Two arrays \curr and \nextA are used to add edges to the current and next sub-levels, respectively.  We also use two boolean arrays \inCurr and \inNextA to mark the edges belonging to \curr and \nextA arrays, respectively.  The \scan procedure is very simple.  To find the edges belonging to an $l$-class, it scans the $S$ array and puts the edges with support equal to $l-2$ in \curr array and marks them in \inCurr.  In a while loop, the edges in array \curr are processed by procedure \procsublvl.  It processes the triangles containing the edges in \curr array. This may decrease the support of the edges of the triangles if the support of the edges in the triangle are higher than $l-2$.  After decreasing the support, if the support of the edges become equal to $l-2$, they are added to \nextA array and they are marked in \inNextA.  Also, an edge in the \curr array is marked as processed in \processedA array once all the triangles containing the edge are processed.  At the end of the procedure \procsublvl, all the edges in \curr are marked as processed and they are also unmarked from the \inCurr array.  The \curr array is swapped with \nextA, \inCurr is swapped with \inNextA, and the processing continues until no more edges can be added to array \nextA.  It is simple to process the edges in serial, since it is performing the same operations as in Algorithm~\ref{alg:serialtruss}.  It processes all the triangles containing an edge with lowest support and mark it processed (delete the edge).  However, to compute \kclass in parallel, the algorithm needs to process and delete the edges in parallel.   

\begin{figure}[!t]
\centering
\includegraphics[width=1.0\columnwidth]{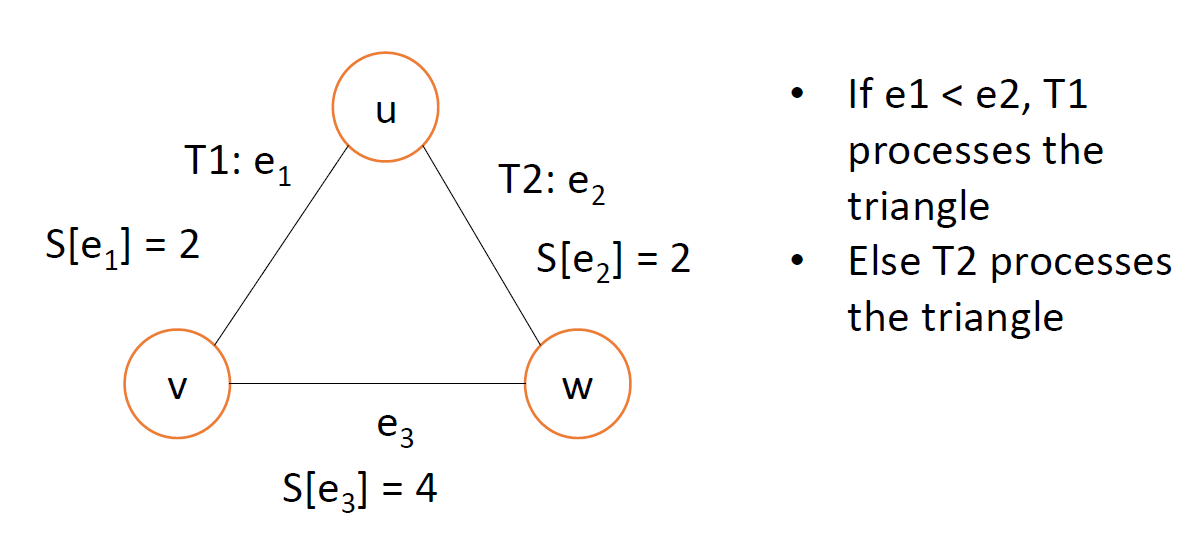}
\caption{Processing of a triangle by threads.  The triangle is processed by the thread containing the lower edge id.}   
\label{fig:triangleOrder}
\end{figure}

Observe that each triangle is processed only once.  This is because the triangles containing an edge are processed when the algorithm computes a \kclass, and the edge is deleted after the processing is done.  So, the triangles do not exist in the graph any more. If a triangle is processed more than once, we can consider the algorithm not to be work-efficient.

\noindent \textbf{Concurrent triangle processing.} Suppose we are computing an $l$-class in parallel. The edges in \curr have support equal to $l-2$ and the edges are processed in parallel in \procsublvl.  Since edges are processed in parallel, the triangles containing the edges are also processed in parallel. This introduces a race condition in processing triangle and updating support of the edges.  For example, let us assume we are processing the triangle $\triangle_{uvw}$ in parallel.  Also assume the edges of the triangles are $e_1 = \langle u, v \rangle$, $e_2 = \langle u, w \rangle$ and $e_3 = \langle v, w \rangle$.  There are three cases to consider when processing this triangle:
\begin{enumerate}
\item only one edge is in \curr
\item two edges are in \curr
\item all the three edges are in \curr
\end{enumerate}
The first case is easy to handle. Since only one edge is in \curr, the thread processing the edge can also process the triangle. The third case is also easy.  Since all the edges are in \curr, the support of all edges is equal to $l-2$ and the triangle could be visited by three different threads. However, the support of the edges will not be updated, as no edge has support greater than $l-2$, and so any thread can process the triangle. The second case is more difficult to handle.  Without loss of generality, assume $e_1$ and $e_2$ have support equal to $l-2$ and $e_1$ is visited by thread $T1$ and $e_2$ is visited by thread $T2$.  Since the $l$-class is computed and the triangle $\triangle_{uvw}$ is processed in this level, and so $S[e_3] > l$. Let us assume $S[e_3] = s$ with $s > l$. Now, $\triangle_{uvw}$ will be processed simultaneously by $T1$ and $T2$ and both threads may find that $S[e_3] > l$ and they will decrease the $S[e_3]$ twice and the support will be changed to $S[e_3] = s - 2$.  However, this is not correct, because each triangle should be processed only once and $S[e_3]$ should be decreased to $s-1$. To solve this problem, we use the ordering of the edges $e_1$ and $e_2$ to enforce processing of a triangle by only one thread.  We decided to process the triangle using the thread that gets the lower edge id.  Thus, $\triangle_{uvw}$ is processed by $T1$ if $e_1 < e_2$ and it is processed by $T2$ if $e_2 < e_1$. The triangle could also be processed by the thread that contains the higher edge id. This helps to process the triangle only once, so $S[e_3]$ would now be correctly decreased to $s-1$. We give a small example in Figure~\ref{fig:triangleOrder}. $S[e_1]$ and $S[e_2]$ are equal to 2 and they are processed by $T1$ and $T2$, respectively.  If $e_1 < e_2$, then the triangle is processed by $T1$ and $S[e_3]$ is decreased to 3 by $T1$. If $e_2 < e_1$, then $S[e_3]$ is decreased to 3 by $T2$.

\begin{algorithm}[!t]
  \caption{Level processing algorithm in \pkt.}
  \label{alg:procsublvl}
  
  \begin{algorithmic}[1]
  	\Procedure{\procsublvl}{\curr, \supA, $l$, \nextA, \inCurr, \inNextA, \processedA, \X}
	\State {Initialize a thread-local array \buff of size $s$} 
	\State {$i \gets 0$} 

	\For{($e_1 \in$ \curr) \textbf{in parallel}} 
		\State $\langle u, v \rangle  \gets e_1$ 
		\For{($j = \mathit{Es}[u]$ to $\mathit{Es}[u+1] - 1$)}
			\State $w \gets N[j]$
			\State $X[w] \gets j +1$
		\EndFor

		\For{($j = \mathit{Es}[v]$ to $\mathit{Es}[v+1] - 1$)}
			\State $w \gets N[j]$

			\If{(\X$[w] = 0 $)}
				\State continue
			\EndIf

			\State $e_2 \gets eid[j]$
			\State $e_3 \gets eid[\mathit{X}[w] -1]$

			\If {(\processedA$[e_2]$ or \processedA$[e_3]$)}
				\State continue
			\EndIf

			\If{(\supA$[e_2] > l $)}  \label{sub:SGreater}
				\If{((($e_1 < e_3$) and  $\mathit{inCurr}[e_3]$) or 
                \Statex~~~~~~~~~~~~~~~~~~~~($\mathit{inCurr}[e_3] = false$)) }

					\State {$a \gets $ atomicSub(\supA$[e_2]$, 1)}

					\If{($a = (l + 1)$)}

						\State {$\mathit{buff}[i] \gets e_2$}; $i \gets i+1$
						\State {$\mathit{inNext}[e_2] \gets true$}

						\If{($i = s$)}
	  						\State Atomically update end of \nextA 	
      							\State {Copy \buff to \nextA}
	  							\State \buff $\gets \phi$; {$i \gets 0$} 		
						\EndIf
					\EndIf

					\If{($a <= l$)}   \label{sub:SLess}
						\State {atomicAdd(\supA$[e_2]$, 1)}
					\EndIf
				\EndIf
			\EndIf

		\EndFor

		\For{($j$ = \numEdges$[u]$ to \numEdges$[u+1] - 1$)}
			\State $w \gets N[j]$
			\State {$\mathit{X}[w] \gets 0$}
		\EndFor

	\EndFor

	\If {($i > 0$)}
		\State Atomically update end of \nextA 	
	  	\State {Copy \buff to \nextA}
	  	\State \buff $\gets \phi$; $i \gets 0$		
	\EndIf

	\For{($e \in$ \curr) \textbf{in parallel}}
		\State {$\mathit{processed}[e] \gets true$}
		\State {$\mathit{inCurr}[e] \gets false$}
	\EndFor

	\EndProcedure

	\end{algorithmic}
\end{algorithm}

The procedure \procsublvl is given in Algorithm~\ref{alg:procsublvl}.  The edges in \curr are processed in parallel.  Assume that an edge $e_1 = \langle u,v \rangle$ is processed by a thread.  To process $e_1$, adjacencies of $u$ are marked in \X.  The adjacencies of $v$ are then visited by the thread and if an adjacency $w$ is marked in $X$, then a triangle is formed by $\triangle_{uvw}$.  Here, $e_2 = \langle v, w \rangle$ and $e_3 = \langle u, w \rangle$.  If any of the edges $e_2$ or $e_3$ is already processed, then the triangle does not exist. If the triangle exists, we mention above that there are three cases to consider. We only show case $(ii)$ in the Algorithm~\ref{alg:procsublvl}.  The thread processes the triangle if $e_1 < e_3$ and $e_3$ is in \curr, or $e_1 < e_2$ and $e_2$ is in \curr array. This is because the thread is processing the edge $e1$.            

\noindent \textbf{Reducing concurrent array additions.} The parallel \scan procedure is given in Algorithm~\ref{alg:pktdescription}.  Since all the threads are adding to the array \curr, the edges need to be added atomically. To decrease the number of atomic operations, each thread uses a buffer \buff and the edges are added to \curr from \buff when the \buff becomes full.  This decreases atomic operations count from $\text{O}(|\mathit{curr}|)$ to $\text{O}(|\mathit{curr}| / |\mathit{buff}|)$. The \scan procedure also marks in array \inCurr the edges that are added to \curr.  Similarly, in procedure \procsublvl, the threads need to use atomic operations to add edges to \nextA array. The number of atomic operations are decreased by assigning a \buff to each thread and copying edges from \buff to \nextA when \buff becomes full.  This also decreases atomic operation counts from  $\text{O}(|\mathit{next}|)$ to $\text{O}(|\mathit{next}| / |\mathit{buff}|)$.  The support of the edges are decreased atomically by the threads.  In Algorithm~\ref{alg:procsublvl}, it can happen that two threads evaluate the condition at Line~\ref{sub:SGreater} true and they decrease the support of an edge below $l-2$.  This is fixed in Line~\ref{sub:SLess}, by atomically increasing support of an edge if it goes below $l-2$.  The procedure \procsublvl marks in array \inNextA the edges that are added to \nextA.  To parallelize \pkt method, the lines from ~\ref{pkt:start} to ~\ref{pkt:end} in Algorithm~\ref{alg:pktdescription} are put in parallel region.  The lines from ~\ref{serial:start} to ~\ref{serial:end} should be executed by a single thread. The algorithm also needs to use synchronization call at the end of \scan procedure, at the end of \procsublvl procedure and after line ~\ref{serial:end}.  Thus, the total number of synchronization calls is \TMax $+ 2S$, where $S = \displaystyle\sum\limits_{i=1}^{t_\text{max}} \mathit{nsl}[i]$ and $\mathit{nsl}[i]$ is the number of sub-levels at level $i$.

\noindent\textbf{Operation counts.} The time complexity for support computation is the same as triangle counting:  $\Theta( \sum_v ( d^+(v) + d^+(v) ^ 2 ) )$. The cumulative time taken by the \scan procedure is $mt_\text{max}$, since the array \supA{ } is of size $m$. The procedure \procsublvl computes the intersection of the end points of each edge exactly once, and so the time complexity is: $\sum_{e=\langle u, v \rangle} ( d(u) + d(v) )$ = $\sum_v d(v)^2$, since we consider an edge $e = \langle u, v \rangle$ and $ u < v$. Since $mt_\text{max} \ll \sum_v d(v)^2$ is small for most real-world graphs, the time complexity is dominated by $\sum_v d(v)^2$. Further, since the wedge count $|\bigwedge| = (\sum_v d(v)^2 - 2m)/2$, and so $|\bigwedge|$ is also an estimate of the work performed.  

Each thread processes the triangles that contains an edge. The number of triangles formed by an edge may vary quite a bit among the edges.  This introduces load imbalance in our algorithm, and we use OpenMP's dynamic loop scheduling to alleviate load imbalance. 


\section{Performance Results and Analysis}
\label{s:results}

In this section, we evaluate the performance of our method \pkt (the source code will be made available at \url{https://github.com/humayunk1}). We also compare our method to our implementations of the \rkt~\cite{Rossi2014} and \WCTruss~\cite{Wang:2012:TDM:2311906.2311909} algorithms. 

\subsection{Experimental Setup}



The methods are evaluated on a dual-socket Intel shared-memory server with 128~GB main memory. The server contains two 2.2 GHz Xeon E5-2650 v4 (Broadwell) processors. Each processor has twelve cores, 30~MB L3 cache, and hyperthreading is turned off. The main memory bandwidth using the STREAM Triad benchmark is 116~GB/s. 

All the codes are compiled using the Intel C/C++ compiler (version 16.0.3) with -O3 optimization. We use OpenMP for parallelization, and threads are pinned to cores using the \emph{compact} pinning strategy.  For the \scan phase, \emph{static} scheduling is used. For support computation and processing edges, \emph{dynamic} scheduling is used with chunk sizes 10 and 4, respectively. 

We choose several large-scale graphs to evaluate our method. These graphs are picked from the University of Florida Sparse Matrix collection~\cite{cite:Davis94} and the Stanford Network Analysis Project~\cite{snapnets} and are listed in Table~\ref{tab:testsuite}. Directed graphs from these sources were made undirected. We also removed self loops and duplicate edges. The graphs are ordered in increasing order of number of wedges ($\bigwedge$), as the wedge count is the closest measure of the amount of work performed by our algorithm. We also list the triangle count, edge count, vertex count, the maximum degree, the maximum coreness, and maximum trussness for each of these graphs. Most of the graphs considered are either snapshots of social networks (soc-pokec, soc-LiveJournal1, ljournal-2008, com-orkut, hollywood-2009, com-friendster) or crawls of web domains (wb-edu, in-2004, uk-2002, indochina-2004, webbcase-2001, arabic-2005, it-2004). We also report the ratio of the number of wedges to the number of triangles. If we use triangle count as an optimistic lower bound for the work performed by a triangle counting (and thus \ktruss decomposition) algorithm, the ratio indicates the possible work reduction that can be achieved if we knew beforehand the edges involved in triangles. as-skitter, for instance, has a very high ratio of 556.89. indochina-2004, on the other hand, has a relatively low ratio of 9.51. Note that the graphs with the highest wedge count, triangle count, edge count, and vertex count are all different. Further, for all the graphs, $c_{\text{max}}$ values are significantly smaller than $d_{\text{max}}$ values.


\begin{table*}[htbp]
  \centering
  \caption{The test suite of graphs used in our study, ordered by number of wedges ($|\bigwedge|$). The number of triangles ($|\triangle|$), vertices ($n$), edges ($m$), maximum degree ($d_{\text{max}}$), maximum coreness ($c_{\text{max}}$), maximum trussness ($t_{\text{max}}$), and wedge-triangle ratio are also given.}
    
    \begin{tabular}{@{}lS[table-format=5.2]S[table-format=2.2]S[table-format=3.2]
    S[table-format=4.2]S[table-format=7]S[table-format=4]S[table-format=4]S[table-format=3.2]@{}}
    \toprule
    
{Graph} & {$|\bigwedge|$ ($\times 10^9$)} & {$|\triangle|$ ($\times 10^9$)}  & {$m$ ($\times 10^6$)} & {$n$ ($\times 10^6$)}  & {$d_{\text{max}}$} & {$c_{\text{max}}$} & {$t_{\text{max}}$} & $\frac{|\bigwedge|}{|\triangle|}$\\

    \midrule
cit-Patents	&	0.34	&	0.01	&	16.52	&	3.77	&	793	&	64	&	36				&	44.68	\\
soc-pokec	&	2.09	&	0.03	&	30.62	&	1.63	&	14854	&	47	&	29				&	64.07	\\
soc-LiveJournal1	&	7.27	&	0.29	&	68.99	&	4.85	&	20333	&	372	&	362				&	25.44	\\
ljournal-2008	&	9.96	&	0.41	&	79.02	&	5.36	&	19432	&	425	&	414				&	24.22	\\
wb-edu	&	12.2	&	0.25	&	57.16	&	9.85	&	25781	&	448	&	449				&	47.91	\\
in-2004	&	15.11	&	0.47	&	16.92	&	1.38	&	21869	&	488	&	489				&	32.54	\\
as-skitter	&	16.02	&	0.03	&	11.1	&	1.7	&	35455	&	111	&	68				&	556.89	\\
com-orkut	&	45.63	&	0.63	&	117.19	&	3.07	&	33313	&	253	&	78				&	72.7	\\
hollywood-2009	&	47.65	&	4.92	&	113.89	&	1.14	&	11467	&	2208	&	2209				&	9.69	\\
uk-2002	&	201.46	&	4.46	&	298.11	&	18.52	&	194955	&	943	&	944				&	45.25	\\
indochina-2004	&	571.94	&	\bfseries	60.12	&	194.11	&	7.41	&	256425	&	\bfseries	6869	&	\bfseries	6851	&	9.51	\\
com-friendster	&	720.65	&	4.17	&	\bfseries	1806.07	&	65.61	&	5214	&	304	&	129			&	172.66	\\
webbase-2001	&	1235.74	&	12.26	&	1019.9	&	\bfseries	118.14	&	816127	&	1506	&	1507			&	100.78	\\
arabic-2005	&	3531.93	&	36.9	&	640	&	22.74	&	575628	&	3247	&	3248				&	95.73	\\
it-2004	&	\bfseries	16163.31	&	48.37	&	1150.73	&	41.29	&	\bfseries	1326744	&	3224	&	3222		&	334.13	\\

     \bottomrule
    \end{tabular}%
  \label{tab:testsuite}%
\end{table*}%

\subsection{Performance Results}

The first step in \pkt is support computation. As discussed in the previous section, our support computation approach is based on a recent triangle counting implementation~\cite{parimalarangan_parsocial2017}. While support computation and triangle counting are related, they are not identical. Triangle counting time can be considered a baseline for support computation, and so we first report parallel triangle counting time in Table~\ref{tab:tricount}. Another baseline is \kcore computation, and we report the time taken by \ourpark~\cite{kabir_parsocial2017} implementation in this table. Further, we demonstrate the performance impact of ordering by reporting triangle counting time using the given (natural) ordering versus reordering vertices in the increasing order of their coreness values. The speedup due to ordering can be as high as $17\times$ (as-skitter). We also give an ordering-dependent algorithm work estimate ($\sum_v d^+(v)^2$). The work estimate ratio serves as an easy-to-compute bound for performance improvement with vertex ordering. Note that this work ratio is 55.7 for as-skitter. We also report $\sum_v d^2(v)$, which is a work estimate for an ordering- and edge orientation-oblivious simple triangle counting implementation. Note that for it-2004, this ratio is as high as 133. This means that if we are considering all wedges and checking if they are closed, the work performed would be nearly two orders of magnitude higher (than our current approach). The use of an efficient algorithm is the reason why the triangle counting time for it-2004 is lower than the time for com-friendster. Because of the considerable impact of ordering on performance, we preprocess all graphs by doing a \kcore decomposition and then reordering vertices. The times for parallel computation of these two steps are also reported in Table~\ref{tab:tricount}. Note that our ordering routine is unoptimized, and so the running time is currently higher than \kcore decomposition time. 

\begin{figure}[!t]
\centering
\includegraphics[width=1.0\columnwidth]{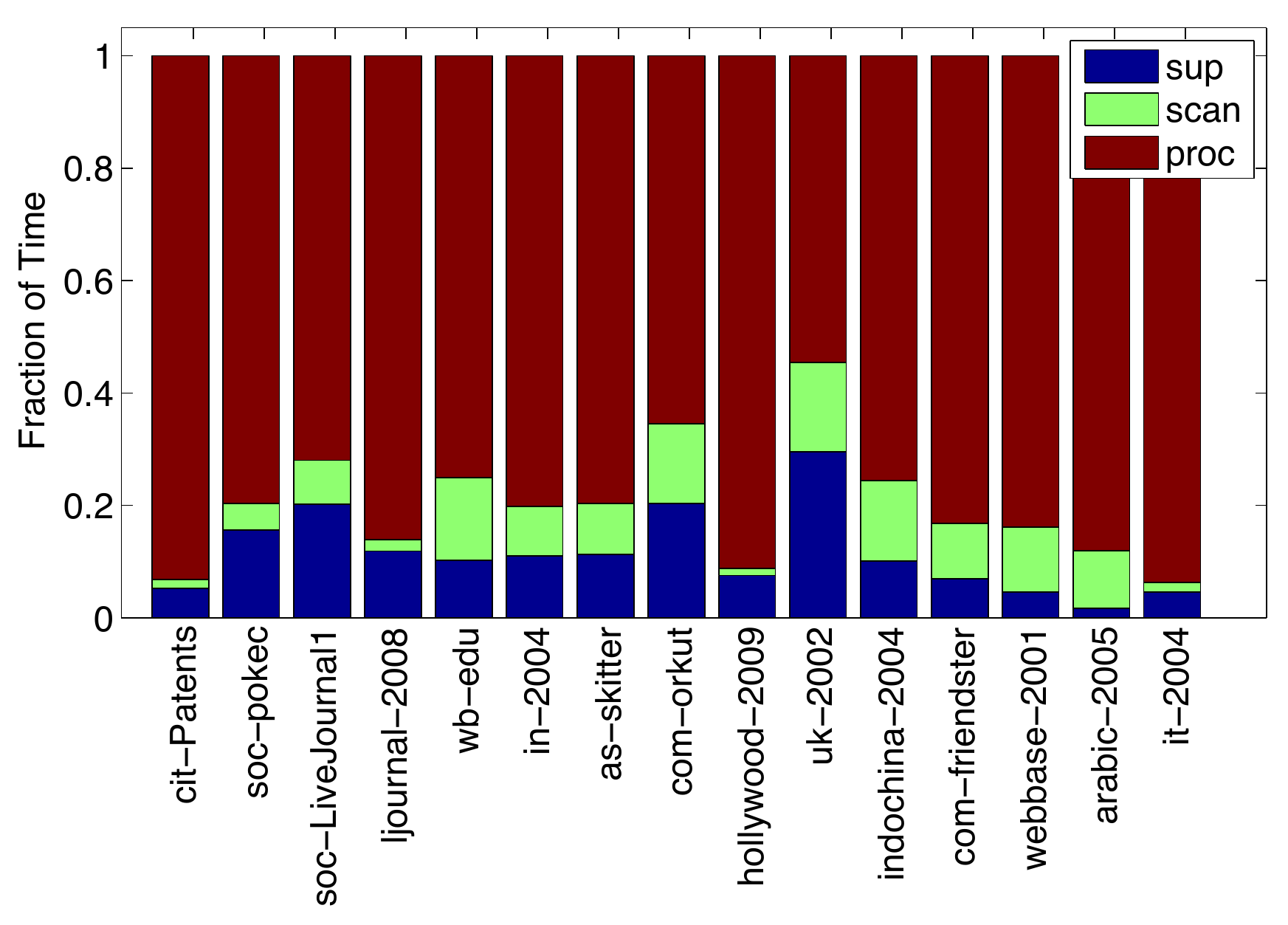}
\caption{Breakdown of \pkt execution time among stages.}   
\label{fig:time_split}
\end{figure}

In Figure~\ref{fig:time_split}, we show the fraction of time spent by our parallel \pkt implementation in each phase (support computation, scan, and edge/wedge processing). The processing phase is consistently the most time-consuming phase for all graphs. For graphs with a relatively high $d_\text{max}$ and edge count, such as uk-2002, the time spent in the scan step is considerable. For graphs where the parallel support computation is more efficient than the naive wedge-counting approaches, the support computation time is lower than processing time. 

We next report \ktruss decomposition time for single-threaded execution of \pkt, and compare the times to single-threaded \rkt and the sequential \WCTruss in Table~\ref{tab:seqtruss}. We also compute a performance rate for \pkt using the wedge count: we report Giga ($10^9$) Wedges processed per second or GWeps. The speedup over \WCTruss gives an indication of the impact of using a hash table (in \WCTruss). Note that our \WCTruss implementation fails to finish in a reasonable amount of time for several graphs, and so we do not report these times. The speedup over single-threaded \rkt is primarily due to faster support computation in \pkt. \pkt's single-threaded GWeps rate ranges from 0.05 to 0.58. The geometric mean of these performance rates is 0.20. Further, the geometric mean of speedup over \rkt is $1.60\times$. The performance rate for social networks is considerably lower than web crawls (e.g., soc-pokec vs. in-2004). Also, the rate for social networks ranges from 0.09 (soc-pokec) to 0.16 (com-friendster), which is a considerably narrower range than the range for web crawls, 0.16 (indochina-2004) to 0.58 (it-2004). It is easier to compare GWeps rates instead of raw execution times.  

\begin{figure}[!t]
\centering  
\includegraphics[width=1.0\columnwidth]{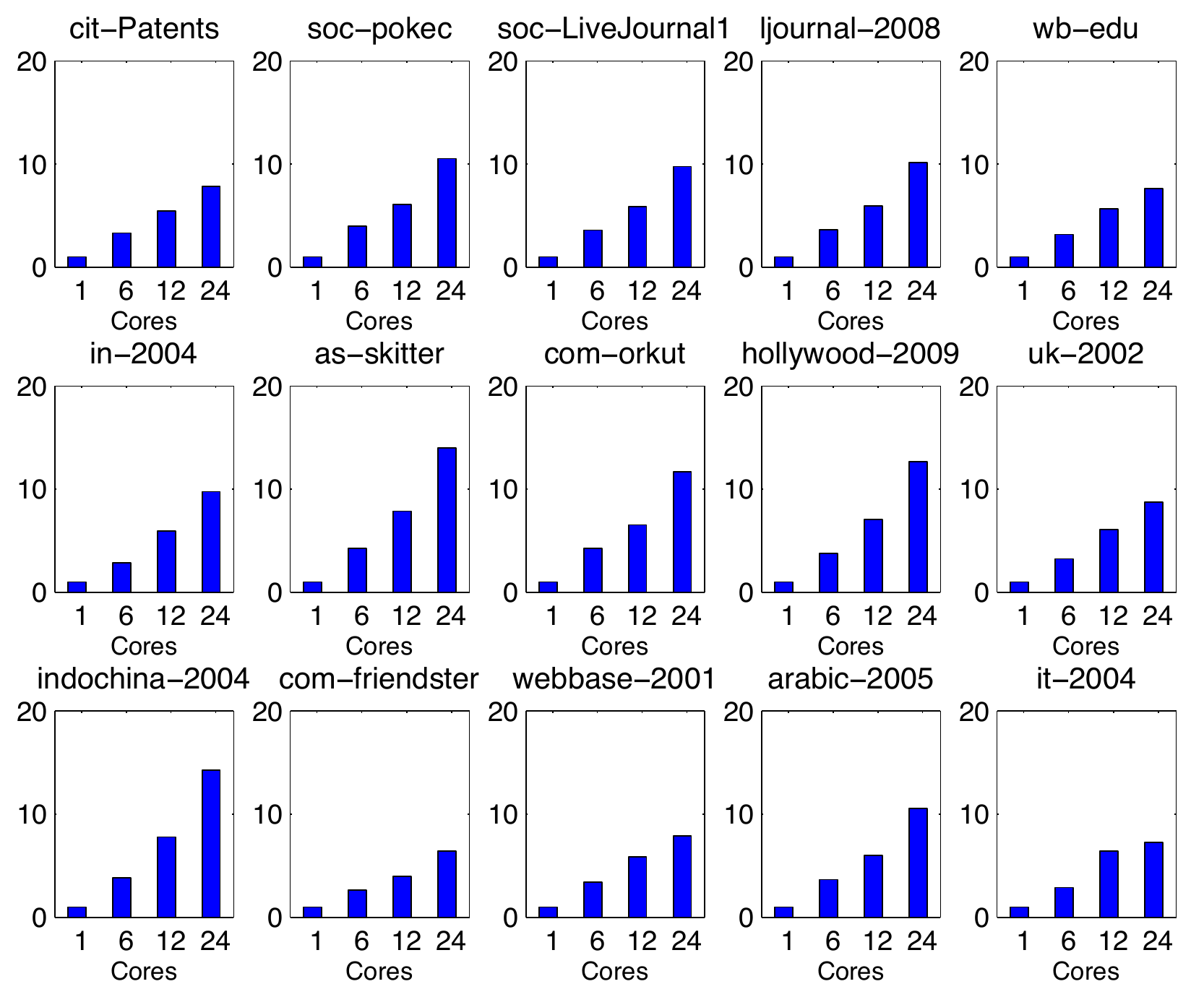}
\caption{\pkt parallel relative scaling.} 
\label{fig:pkt_speedup}
\end{figure}

We next report relative parallel speedup for multithreaded execution in Figure~\ref{fig:pkt_speedup}. We observe reasonable scaling within a socket, as well as performance improvement going from one to two sockets. Table~\ref{tab:partruss} gives \pkt 24-core execution times and rates. Additionally, we report speedup over parallel \rkt. The 24-core relative speedup (geometric mean) is $9.68\times$. Since only the support computation phase in \rkt is parallel, we observe a considerable speedup overall (geometric mean  $12.94\times$). Further, the multithreaded performance rate ranges from 0.42 to 6.45, with a geometric mean of 1.93. Parallel \rkt failed to finish for it-2004, and so we do not report this speedup. We again note that the performance rates for web crawls are typically higher than the rates for social networks. This is partially due to better multithreaded scaling for web crawls (e.g., 14.29$\times$ for indochina-2004 vs. 6.41$\times$ for com-friendster).

\begin{figure}[!t]
\begin{center}  
\includegraphics[width=1.0\columnwidth]{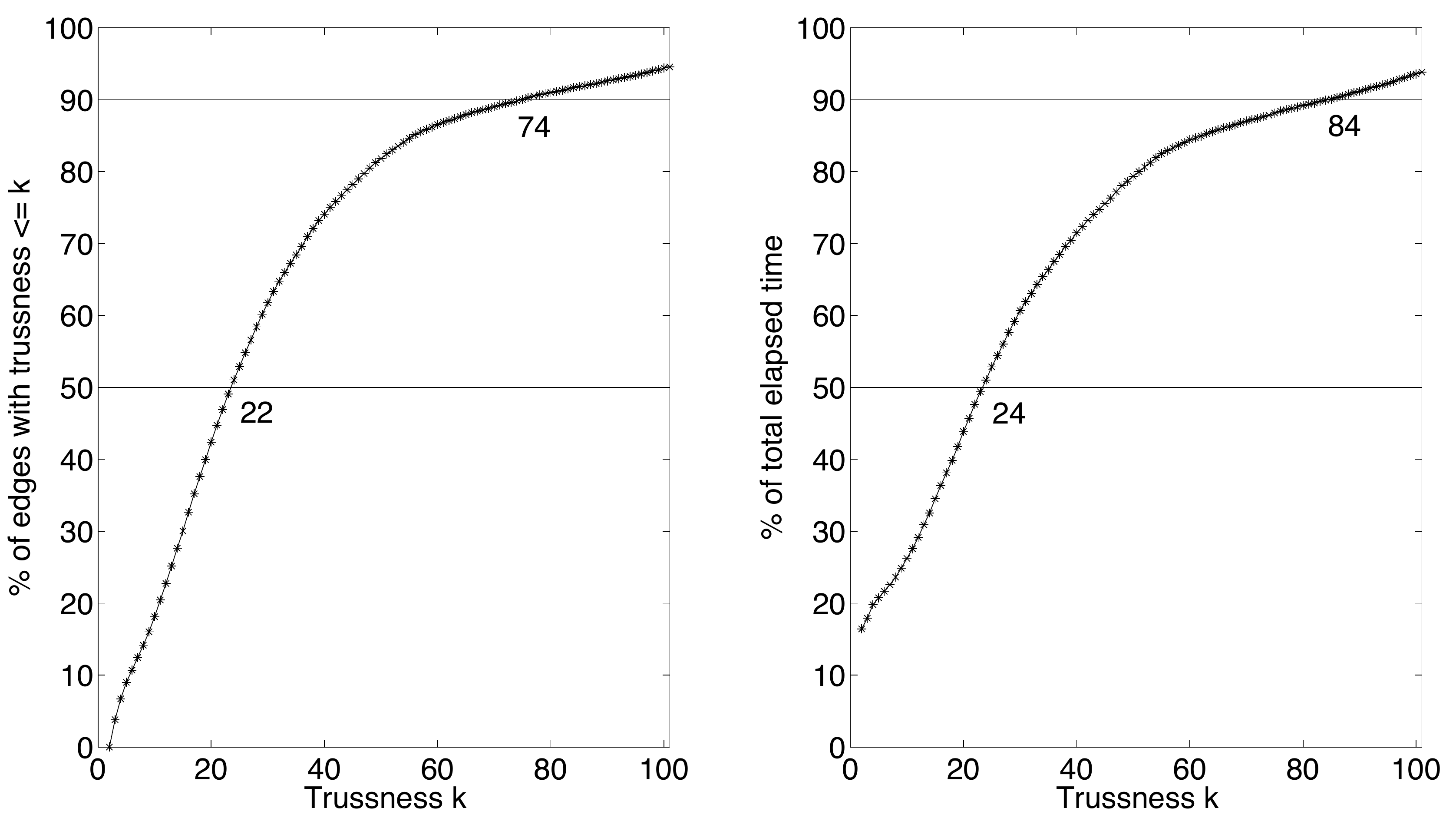}
\end{center}
\caption{Trussness and execution time distributions for uk-2002. 50\% of edges have trussness less than 22 and 90\% of edges have trussness less than 74. Similarly, 50\% of total time (24-core execution) is spent processing edges of trussness less than 24, and 90\% of total time is spent processing edges with trussness less than 84. Note that $t_{max}$ for this graph is 944, but we only show results up to trussness 100.}   
\label{fig:vtxineachlevel}
\end{figure}

Finally, we show in Figure~\ref{fig:vtxineachlevel} that the parallel performance is closely correlated with the wedge count work estimate, and that the maximum trussness value does not have a considerable impact on performance. Our parallelization can be considered work-efficient in the sense that we do not incur a significant overhead for barrier synchronization introduced due to a high $t_\text{max}$ observed in real-world graphs.

Our single-node performance rates are better than the rates reported by Wang et al.~\cite{Wang:2012:TDM:2311906.2311909} and Chen et al.~\cite{7004264}. For instance, Wang~et~al.\ report a sequential as-skitter time of 281 seconds, whereas our serial time is 34.71 seconds (8.08$\times$ faster). In~\cite{7004264}, the reported time for the graph com-dblp is more than 50 seconds with 12 reducers, while our serial algorithm takes just 0.4 seconds for this graph. Further, \pkt is faster than the best shared-memory results reported in~\cite{SSP17}. For the graphs soc-LiveJournal1, as-skitter, and com-orkut,~\cite{SSP17} reports execution times of  104.6, 13.8, and 359.1 seconds, respectively, on a 24-core Intel server with Ivy Bridge processors. For the same graphs, \pkt takes 7.35, 2.48, and 26.27 seconds, respectively, on our 24-core Broadwell system. Thus, for these graphs, \pkt is 5.56$\times$ to 14.23$\times$ faster than~\cite{SSP17} (ignoring the difference in hardware).

\begin{table*}[htbp]
  \centering
  \caption{Impact of vertex ordering on triangle counting time: parallel triangle counting time with increasing \kcore order (KCO) and natural ordering (NAT). Triangle counting work estimates are also given. Parallel \kcore computation time and \kcore reordering times are also given.}
    \begin{tabular}{@{}lS[table-format=2.2]S[table-format=3.2]S[table-format=2.3]S[table-format=3.2]S[table-format=4.2]S[table-format=2.2] S[table-format=5.2] S[table-format=3.2] S[table-format=2.2] S[table-format=3.2] @{}}
    \toprule
              & \multicolumn{2}{c}{$\triangle$ time (s)} & {KCO} & \multicolumn{2}{c}{$\sum_{v \in V} (d^+(v)^2) (\times 10^9)$} & {Work} &    {$\sum d(v)^2$}   &  {$\sum d(v)^2/$}     & {\kcore}      & {Ordering} \\    
    Graph & {KCO}   & {NAT}   & {Speedup} & {KCO}  & {NAT}  & {Ratio} & {($\times 10^9$)} & {$\sum d^+(v)^2$} & {time (s)} & {time (s)} \\
    \cmidrule{1-1} \cmidrule(lr){2-3} \cmidrule{4-4} \cmidrule(lr){5-6} \cmidrule{7-7} \cmidrule(lr){8-9} \cmidrule{10-11}
    cit-Patents & 0.12  & 0.13  & 1.08  & 0.24  & 0.32  & 1.37  & 0.70  & 2.99  & 0.08  & 0.34 \\
    soc-pokec & 0.24  & 0.20  & 0.85  & 1.20  & 2.80  & 2.35  & 4.22  & 3.53  & 0.11  & 0.59 \\
    soc-LiveJournal1 & 0.47  & 0.60  & 1.27  & 3.06  & 10.97 & 3.59  & 14.62 & 4.78  & 0.26  & 1.94 \\
    ljournal-2008 & 0.55  & 0.31  & \bfseries 0.57  & 4.16  & 6.79  & 1.63  & 20.02 & 4.81  & 0.25  & 2.34 \\
    wb-edu & 0.20  & 0.43  & 2.09  & 1.38  & 11.58 & 8.38  & 24.50 & 17.73 & 0.12  & 1.15 \\
    in-2004 & 0.11  & 0.42  & 3.86  & 1.64  & 16.31 & 9.97  & 30.24 & 18.49 & 0.07  & 0.40 \\
	as-skitter & 0.07  & 1.11  & \bfseries 17.00 & 0.56  & 31.17 & \bfseries 55.70 & 32.07 & 57.31 & 0.07  & 0.60 \\
      com-orkut & 2.14  & 3.10  & 1.45  & 19.36 & 68.08 & 3.52  & 91.49 & 4.72  & 0.58  & 4.12 \\
    hollywood-2009 & 1.46  & 1.02  & 0.70  & 29.71 & 43.35 & 1.46  & 95.40 & 3.21  & 0.30  & 3.24 \\
      uk-2002 & 1.44  & 4.50  & 3.12  & 18.09 & 187.73 & 10.38 & 403.44 & 22.30 & 0.49  & 5.80 \\
    indochina-2004 & 9.21  & 11.38 & 1.24  & 184.16 & 485.10 & 2.63  & 1144.18 & 6.21  & 0.80  & 4.68 \\
    com-friendster & \bfseries 61.02 & 165.57 & 2.71  & 398.64 & 815.01 & 2.04  & 1444.92 & \bfseries 3.62  & \bfseries 21.23 & \bfseries 112.59 \\
    webbase-2001 & 4.94  & 14.90 & 3.02  & 52.22 & 546.91 & 10.47 & 2473.20 & 47.36 & 2.07  & 21.60 \\
	arabic-2005 & 7.24  & 62.17 & 8.59  & 134.47 & 3155.83 & 23.47 & 7064.97 & 52.54 & 1.58  & 15.82 \\
    it-2004 & \bfseries 12.91 & 71.99 & 5.58  & 242.98 & 4482.33 & 18.45 & 32328.67 & \bfseries 133.05 & 3.39  & 31.10 \\
    \bottomrule
    \end{tabular}%
  \label{tab:tricount}%
\end{table*}%

\begin{table}[htbp]
\setlength{\tabcolsep}{3pt}
\centering
  \caption{\ktruss decomposition sequential algorithm performance. Empty cells indicate that the implementation did not finish in 1 hour.}
    \begin{tabular}{@{}lS[table-format=5.2]S[table-format=4.2]S[table-format=5.2]S[table-format=1.2]S[table-format=1.2]@{}}
    \toprule
      & \multicolumn{3}{c}{Execution time (s)} & {Perf} & {Speedup}\\
Graph &  {PKT} & {\WCTruss} & {\rkt} & {(GWeps)} & {over \rkt}\\
    \cmidrule{1-1} \cmidrule(lr){2-4} \cmidrule(r){5-5} \cmidrule{6-6}
cit-Patents	&	6.29	&	46.17	&	8.16	&	0.05	&	1.3	\\
soc-pokec	&	22.54	&	1037.76	&	35.32	&	0.09	&	1.57	\\
soc-LiveJournal1	&	71.74	&	4175.15	&	110.25	&	0.1	&	1.54	\\
ljournal-2008	&	94.47	&	6832.99	&	147.28	&	0.11	&	1.56	\\
wb-edu	&	44.01	&	620.92	&	69.79	&	0.28	&	1.59	\\
in-2004	&	33.35	&	1676.91	&	54.86	&	0.45	&	1.64	\\
as-skitter	&	34.71	&	339.38	&	70.36	&	0.46	&	2.03	\\
com-orkut	&	423.76	&	{--}	&	688.81	&	0.11	&	1.63	\\
hollywood-2009	&	434.54	&	{--}	&	676.44	&	0.11	&	1.56	\\
uk-2002	&	580.23	&	{--}	&	1130.81	&	0.35	&	1.95	\\
indochina-2004	&	3547	&	{--}	&	3155.22	&	0.16	&	0.89	\\
com-friendster	&	4597.59	&	{--}	&	{--}	&	0.16	&	{--}	\\
webbase-2001	&	3372.45	&	{--}	&	6746.66	&	0.37	&	2	\\
arabic-2005	&	7335.33	&	{--}	&	14745.7	&	0.48	&	2.01	\\
it-2004	&	27658.28	&	{--}	&	{--}	&	0.58	&	{--}	\\
    \bottomrule
    \end{tabular}%
  \label{tab:seqtruss}%
\end{table}%

\begin{table}[htbp]
\setlength{\tabcolsep}{3pt}
  \centering
  \caption{\ktruss decomposition PKT parallel performance.}
    \begin{tabular}{@{}lS[table-format=4.2]S[table-format=1.2]S[table-format=2.2]S[table-format=2.2]@{}}
    \toprule
         & {Time} & {Perf}    & \multicolumn{2}{c}{Speedup}\\
   Graph & {(s)}  & {(GWeps)} & {24-core} & {over \rkt}\\
    \cmidrule{1-1} \cmidrule(lr){2-2} \cmidrule{3-3} \cmidrule(l){4-5}
cit-Patents	&	0.8	&	0.42	&	7.86	&	8.51	\\
soc-pokec	&	2.14	&	0.97	&	10.53	&	13.68	\\
soc-LiveJournal1	&	7.35	&	0.99	&	9.77	&	12.38	\\
ljournal-2008	&	9.3	&	1.07	&	10.16	&	12.91	\\
wb-edu	&	5.77	&	2.11	&	7.63	&	9.42	\\
in-2004	&	3.42	&	4.42	&	9.77	&	10.83	\\
as-skitter	&	2.48	&	6.45	&	13.97	&	17.99	\\
com-orkut	&	36.27	&	1.26	&	11.68	&	15.26	\\
hollywood-2009	&	34.36	&	1.39	&	12.65	&	17	\\
uk-2002	&	66.52	&	3.03	&	8.72	&	13.06	\\
indochina-2004	&	248.15	&	2.3	&	14.29	&	8.97	\\
com-friendster	&	1660.5	&	1	&	6.41	&	22.56	\\
webbase-2001	&	427.8	&	2.89	&	7.88	&	12.06	\\
arabic-2005	&	693.51	&	5.09	&	10.58	&	13	\\
it-2004	&	3817.39	&	4.23	&	7.25	&	{--}	\\

    \bottomrule
    \end{tabular}%
  \label{tab:partruss}%
\end{table}%

\section{Conclusions and Future Work}
\label{s:conclusions}

We presented a new algorithm \pkt for parallel truss decomposition on shared-memory multicore platforms. The algorithm has several novel ideas, including a strategy to avoid the inherently-sequential edge processing, an optimized edge triangle support estimation method, and a choice of data structures designed to reduce memory use and parallelization overhead. On a 24-core system, we demonstrate a mean parallel speedup of $9.68\times$ for a collection of real-world graph instances. Future work related to the presented algorithm includes strategies to improve load balance in the edge processing phase and further reduce memory use. Also, porting this algorithm to GPU and distributed-memory settings appears to be non-trivial.   

\section*{Acknowledgments}

This work is supported by the US National Science Foundation grants ACI-1253881 and CCF-1439057.

\bibliographystyle{IEEEtran}
\bibliography{IEEEabrv,refs}

\begin{thebibliography}{10}
\providecommand{\url}[1]{#1}
\csname url@samestyle\endcsname
\providecommand{\newblock}{\relax}
\providecommand{\bibinfo}[2]{#2}
\providecommand{\BIBentrySTDinterwordspacing}{\spaceskip=0pt\relax}
\providecommand{\BIBentryALTinterwordstretchfactor}{4}
\providecommand{\BIBentryALTinterwordspacing}{\spaceskip=\fontdimen2\font plus
\BIBentryALTinterwordstretchfactor\fontdimen3\font minus
  \fontdimen4\font\relax}
\providecommand{\BIBforeignlanguage}[2]{{%
\expandafter\ifx\csname l@#1\endcsname\relax
\typeout{** WARNING: IEEEtran.bst: No hyphenation pattern has been}%
\typeout{** loaded for the language `#1'. Using the pattern for}%
\typeout{** the default language instead.}%
\else
\language=\csname l@#1\endcsname
\fi
#2}}
\providecommand{\BIBdecl}{\relax}
\BIBdecl

\bibitem{Luce1950}
R.~D. Luce, ``Connectivity and generalized cliques in sociometric group
  structure,'' \emph{Psychometrika}, vol.~15, no.~2, pp. 169--190, 1950.

\bibitem{Mokken:1979}
R.~J. Mokken, ``{Cliques, clubs and clans},'' \emph{Quality \& Quantity},
  vol.~13, pp. 161--173, Apr. 1979.

\bibitem{doi:10.1080/0022250X.1978.9989883}
S.~B. Seidman and B.~L. Foster, ``A graph-theoretic generalization of the
  clique concept,'' \emph{J.\ Mathematical Sociology}, vol.~6, no.~1, pp.
  139--154, 1978.

\bibitem{Abello2002}
J.~Abello, M.~G.~C. Resende, and S.~Sudarsky, ``Massive quasi-clique
  detection,'' in \emph{Proc.\ Latin American Theoretical Informatics Symp.\
  (LATIN)}, 2002.

\bibitem{Pei:2005:MCQ:1081870.1081898}
J.~Pei, D.~Jiang, and A.~Zhang, ``On mining cross-graph quasi-cliques,'' in
  \emph{Proc.\ Int'l.\ Conf.\ on Knowledge Discovery in Data Mining (KDD)},
  2005.

\bibitem{Sei83}
S.~B. Seidman, ``Network structure and minimum degree,'' \emph{Social
  Networks}, vol.~5, no.~3, pp. 269--287, 1983.

\bibitem{MB83}
D.~Matula and L.~Beck, ``Smallest-last ordering and clustering and graph
  coloring algorithms,'' \emph{J.\ ACM}, vol.~30, no.~3, pp. 417--427, 1983.

\bibitem{CohenTruss}
J.~Cohen, ``Trusses: Cohesive subgraphs for social network analysis,'' National
  Security Agency, Tech. Rep., 2008.

\bibitem{SYK06}
K.~Saito, T.~Yamada, and K.~Kazama, ``Extracting communities from complex
  networks by the $k$-dense method,'' in \emph{Proc.\ Int'l.\ Workshop on
  Mining Complex Data (MCD)}, 2006.

\bibitem{6228155}
Y.~Zhang and S.~Parthasarathy, ``Extracting analyzing and visualizing triangle
  k-core motifs within networks,'' in \emph{Proc.\ Int'l.\ Conf.\ on Data
  Engineering (ICDE)}, 2012.

\bibitem{VB13}
A.~Verma and S.~Butenko, ``Network clustering via clique relaxations: A
  community based approach,'' in \emph{Graph Partitioning and Graph
  Clustering}, D.~Bader, H.~Meyerhenke, P.~Sanders, and D.~Wagner, Eds.\hskip
  1em plus 0.5em minus 0.4em\relax AMS, 2013, ch.~9, pp. 129--139.

\bibitem{SSPC15}
A.~E. Sar{\i}y{\"u}ce, C.~Seshadhri, A.~Pinar, and {\"U}.~V.
  {\c{C}}ataly{\"u}rek, ``Finding the hierarchy of dense subgraphs using
  nucleus decompositions,'' in \emph{Proc.\ Int'l.\ Conf.\ on World Wide Web
  (WWW)}, 2015.

\bibitem{DBLP:conf/nips/Alvarez-HamelinDBV05}
J.~I. Alvarez{-}Hamelin, L.~Dall'Asta, A.~Barrat, and A.~Vespignani, ``Large
  scale networks fingerprinting and visualization using the k-core
  decomposition,'' in \emph{Proc.\ Advances in Neural Information Processing
  Systems (NIPS)}, 2005.

\bibitem{Huang:2015:ACC:2856318.2856323}
X.~Huang, L.~V.~S. Lakshmanan, J.~X. Yu, and H.~Cheng, ``Approximate closest
  community search in networks,'' \emph{Proc. VLDB Endow.}, vol.~9, no.~4, pp.
  276--287, 2015.

\bibitem{DBLP:journals/corr/abs-1302-6256}
R.~A. Rossi, D.~F. Gleich, and A.~H. Gebremedhin, ``Parallel maximum clique
  algorithms with applications to network analysis,'' \emph{SIAM Journal on
  Scientific Computing}, vol.~37, no.~5, pp. C589--C616, 2015.

\bibitem{Wang:2012:TDM:2311906.2311909}
J.~Wang and J.~Cheng, ``Truss decomposition in massive networks,'' \emph{Proc.
  VLDB Endow.}, vol.~5, no.~9, pp. 812--823, May 2012.

\bibitem{5076317}
J.~Cohen, ``Graph twiddling in a {MapReduce} world,'' \emph{Computing in
  Science \& Engineering}, vol.~11, no.~4, pp. 29--41, 2009.

\bibitem{7004264}
P.~L. Chen, C.~K. Chou, and M.~S. Chen, ``Distributed algorithms for k-truss
  decomposition,'' in \emph{Proc.\ Int'l.\ Conf.\ on Big Data (Big Data)},
  2014.

\bibitem{SSP17}
A.~E. Sar{\i}y{\"u}ce, C.~Seshadhri, and A.~Pinar, ``Parallel local algorithms
  for core, truss, and nucleus decompositions,'' arXiv.org e-Print archive,
  \url{https://arxiv.org/abs/1704.00386}, 2017.

\bibitem{Graphulo}
V.~Gadepally, J.~Bolewski, D.~Hook, D.~Hutchison, B.~Miller, and J.~Kepner,
  ``Graphulo: Linear algebra graph kernels for {NoSQL} databases,'' in
  \emph{Proc.\ Workshop on Graph Algorithms Building Blocks (GABB)}, 2015.

\bibitem{GraphChallenge}
S.~Samsi, V.~Gadepally, M.~Hurley, M.~Jones, E.~Kao, S.~Mohindra,
  P.~Monticciolo, A.~Reuther, S.~Smith, W.~Song, D.~Staheli, and J.~Kepner,
  ``Static graph challenge: Subgraph isomorphism,'' MIT Lincoln Laboratory,
  Tech. Rep., 2017,
  \url{http://graphchallenge.mit.edu/sites/default/files/documents/SubgraphIsomorphismChallenge-2017-06-15.pdf}.

\bibitem{7004366}
N.~S. Dasari, D.~Ranjan, and M.~Zubair, ``{ParK}: An efficient algorithm for
  k-core decomposition on multicore processors,'' in \emph{Proc.\ Int'l.\
  Workshop on High Performance Big Graph Data Management, Analysis, and Mining
  (BigGraphs)}, 2014.

\bibitem{DBLP:journals/corr/cs-DS-0310049}
V.~Batagelj and M.~Zaversnik, ``An {$O(m)$} algorithm for cores decomposition
  of networks,'' arXiv.org e-Print archive,
  \url{http://arxiv.org/abs/cs.DS/0310049}, 2003.

\bibitem{Rossi2014}
R.~A. Rossi, ``Fast triangle core decomposition for mining large graphs,'' in
  \emph{Proc.\ Pacific-Asia Conf.\ on Advances in Knowledge Discovery and Data
  Mining (PAKDD)}, 2014.

\bibitem{Malewicz:2010:PSL:1807167.1807184}
G.~Malewicz, M.~H. Austern, A.~J. Bik, J.~C. Dehnert, I.~Horn, N.~Leiser, and
  G.~Czajkowski, ``Pregel: A system for large-scale graph processing,'' in
  \emph{Proc.\ Int'l.\ Conf.\ on Management of Data (SIGMOD)}, 2010.

\bibitem{CN85}
N.~Chiba and T.~Nishizeki, ``Arboricity and subgraph listing algorithms,''
  \emph{SIAM J. Comput.}, vol.~14, no.~1, pp. 210--223, 1985.

\bibitem{Lat08}
M.~Latapy, ``Main-memory triangle computations for very large (sparse
  (power-law)) graphs,'' \emph{Theoretical Computer Science}, vol. 407, no.
  1--3, pp. 458--473, 2008.

\bibitem{OB14}
M.~Ortmann and U.~Brandes, ``Triangle listing algorithms: Back from the
  diversion,'' in \emph{Proc.\ Workshop on Algorithm Engineering and
  Experiments (ALENEX)}, 2014.

\bibitem{shun7113280}
J.~Shun and K.~Tangwongsan, ``Multicore triangle computations without tuning,''
  in \emph{Proc.\ Int'l.\ Conf.\ on Data Engineering (ICDE)}, 2015, pp.
  149--160.

\bibitem{CXL16}
Y.~Cui, D.~Xiao, and D.~Loguinov, ``On efficient external-memory triangle
  listing,'' in \emph{Proc.\ Int'l.\ Conf.\ on Data Mining (ICDM)}, 2016.

\bibitem{Xiao:2017:ACT:3034786.3034790}
D.~Xiao, Y.~Cui, D.~B. Cline, and D.~Loguinov, ``On asymptotic cost of triangle
  listing in random graphs,'' in \emph{Proc.\ Symp.\ on Principles of Database
  Systems (PODS)}, 2017.

\bibitem{parimalarangan_parsocial2017}
S.~Parimalarangan, G.~M. Slota, and K.~Madduri, ``Fast parallel graph triad
  census and triangle counting on shared-memory platforms,'' in \emph{Proc.\
  Workshop on Parallel and Distributed Processing for Computational Social
  Systems {(ParSocial)}}, 2017.

\bibitem{kabir_parsocial2017}
H.~Kabir and K.~Madduri, ``Parallel k-core decomposition on multicore
  platforms,'' in \emph{Proc.\ Workshop on Parallel and Distributed Processing
  for Computational Social Systems {(ParSocial)}}, 2017.

\bibitem{DBLP:journals/corr/abs-1103-5320}
A.~Montresor, F.~De~Pellegrini, and D.~Miorandi, ``Distributed k-core
  decomposition,'' in \emph{Proc.\ Symp.\ on Principles of Distributed
  Computing (PODC)}, 2011.

\bibitem{cite:Davis94}
T.~A. Davis and Y.~Hu, ``The {University} of {Florida} sparse matrix
  collection,'' \emph{ACM Transactions on Mathematical Software}, vol.~38, pp.
  1:1--1:25, 2011, \url{http://www.cise.ufl.edu/research/sparse/matrices}.

\bibitem{snapnets}
J.~Leskovec and A.~Krevl, ``{SNAP Datasets}: {Stanford} large network dataset
  collection,'' \url{http://snap.stanford.edu/data}, Jun. 2014.

\end{thebibliography}

\end{document}